\documentclass[aps,prl,reprint,superscriptaddress,amsmath,amssymb,longbibliography,floatfix,footinbib,nobibnotes]{revtex4-2}

\usepackage{graphicx}
\usepackage{dcolumn}
\usepackage{bm}
\usepackage{xcolor}
\usepackage{epsf}
\usepackage{epsfig}

\begin{document}

\title{Angular dependence of the Wigner time delay upon tunnel ionization of CO}

\author{N. Anders}
\email{anders@atom.uni-frankfurt.de}
\author{A. Geyer}
\author{M. Hofmann}
\author{D. Trabert}
\author{J. Stindl}
\author{N. L. Wong}
\author{J. Kruse}
\author{S. Jacob}
\author{M. S. Schöffler}
\author{L. Ph. H. Schmidt}
\affiliation{Institut f\"ur Kernphysik, Goethe-Universit\"at, Max-von-Laue-Str. 1, 60438 Frankfurt am Main, Germany}

\author{T. Jahnke}
\affiliation{Max-Planck-Institut f\"ur Kernphysik, Saupfercheckweg 5, 69117 Heidelberg, Germany}

\author{M. Kunitski}
\author{S. Eckart}
\email{eckart@atom.uni-frankfurt.de}
\affiliation{Institut f\"ur Kernphysik, Goethe-Universit\"at, Max-von-Laue-Str. 1, 60438 Frankfurt am Main, Germany}

\date{\today}
\begin{abstract}
We report on Wigner time delays in the strong-field ionization of carbon monoxide. Using co-rotating two-color laser fields (395 and 790 nm), we retrieve time delays that depend on the electron emission direction in the molecular frame, with variations of up to $\pm70$\nobreak\,as. The measured angular dependence agrees with predictions from a simple model based on molecular Ammosov–Delone–Krainov (MO-ADK) theory. Our results demonstrate that the directional dependence of electron emission times in asymmetric molecular potentials can be experimentally accessed in the strong-field regime.
\end{abstract}

\maketitle
The advent of attosecond technology has opened the route to address the photoelectric effect also in the time domain. To quantify time scales in photoionization, the Wigner time delay is a well-established concept \cite{Wigner1954}. It is defined as $\tau_{\mathrm{W}} = \hbar \frac{d}{dE} \arg(\psi)$, where $E$ is the kinetic energy of the electron and $\arg(\psi)$ is the phase of the continuum photoelectron wave function $\psi$. For atoms, Wigner time delays have been extensively investigated \cite{Schultze2010,Klunder2011,Gruson2016,Heuser2016}. In molecular systems, the Wigner time delay $\tau_{\mathrm{W}}$ is an even more interesting quantity, as it generally depends on the electron emission direction $\beta$ relative to the molecular axis in the plane of polarization. $\tau_{\mathrm{W}}$ can be decomposed into an angle-dependent contribution and an angle-independent contribution. Thus, for molecules, the Wigner time delay can be written as $\tau_{\mathrm{W}} = \tau_\mathrm{W,A}(E)+\Delta\tau_\mathrm{W,M}(E,\beta)$. Here, $\Delta\tau_\mathrm{W,M}(E,\beta)$ accounts for delays that are due to non-centrosymmetric molecular potentials or non-isotropic electronic states. $\tau_\mathrm{W,A}(E)$ is the angle-independent contribution to the Wigner time delay, which is also present in atoms.

Wigner time delays have been investigated extensively in the single-photon regime \cite{vos2018orientation,Gong2022} using reconstruction of attosecond harmonic beating by interference of two-photon transitions (RABBITT) or wave packet extraction methods employing synchrotron radiation \cite{Rist2021, Holzmeier2021} to gain insights into molecular photoionization. Recently, also in the strong-field regime Wigner time delays became accessible using holographic angular streaking of electrons (HASE) \cite{Han2018,Eckart_HASEtheo,Trabert2021,ArgonHASE,Trabert2023,Guo2022}, which can be considered the strong-field analogue of RABBITT. HASE uses a sub-sycle interference pattern (the alternative half-ring pattern) to access properties of the phase of the electronic wavefunction in momentum representation. Previous work relying on HASE demonstrated that in molecular hydrogen the Wigner time delay varies by up to $\pm 40$\nobreak\,as as a function of the electron emission direction in the molecular frame \cite{Trabert2021} and even by up to $\pm 200$\nobreak\,as for argon dimers \cite{ArgonHASE}. Further, it was shown that these time delays can be related to microscopic, sub-nanometer position offsets of the electronic wave packet at the tunnel exit \cite{Eckart_HASEtheo}. More recently, Guo et al. calculated Wigner time delays for carbon monoxide (CO) in the strong-field regime using a molecular quantum trajectory Monte Carlo model and numerical solutions of the time-dependent Schrödinger equation. They also carried out a proof-of-concept experiment obtaining molecular-frame photoelectron angular emission distributions, while the corresponding photoemission delays were retrieved employing the aforementioned modeling \cite{Guo2022}. However, a direct experimental determination of Wigner time delays of a heteronuclear molecule in the strong-field regime is still lacking.

In the present Letter, we report on the direct experimental measurement of the molecular-frame Wigner time delays $\Delta\tau_\mathrm{W,M}(E,\beta)$ for the asymmetric CO molecule. To this end, we use HASE, which is an interferometric technique that builds on a sub-cycle interference in momentum space \cite{Eckart2020_sideband}. From the phase of the electronic wave function in momentum space, we retrieve position offsets at the tunnel exit and, eventually, Wigner time delays.

\begin{figure}[t]
\includegraphics[width=\columnwidth]{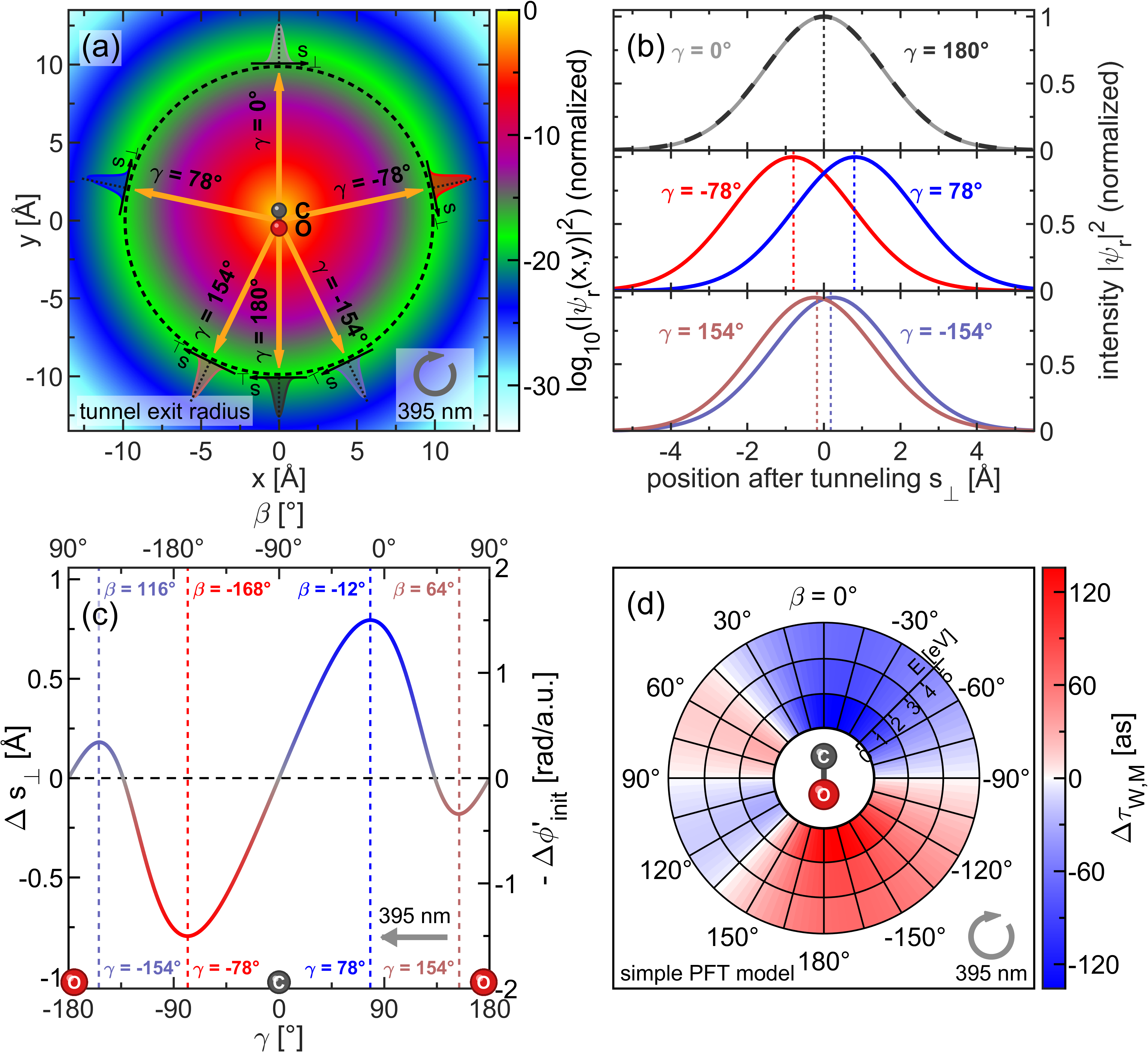}
\caption{\label{fig1} Origin of the angular-dependent Wigner time delay in the strong-field ionization of carbon monoxide (CO). (a)~Illustration of the bound electronic orbital of CO in position space that is relevant for tunnel ionization. Exemplary tunneling directions (defined by the angle $\gamma$, which denotes the tunneling direction with respect to the molecular axis in the plane of polarization) are indicated: the parallel alignments ($0^\circ, 180^\circ$) and the angles of maximum position offset ($\gamma = \pm 78^\circ, \pm 154^\circ$). The colored curves along the transverse axes $s_{\perp}$ relative to the tunneling direction visualize the local one-dimensional projection of the electron's density distribution at the tunnel exit. (b)~Detailed view of these 1D projections along $s_{\perp}$. The asymmetric shape of the molecular orbital induces a $\gamma$-dependent position offset $\Delta s_{\perp}$. (c)~The position offset $\Delta s_{\perp}(\gamma)$ (left axis) is equivalent to a phase gradient $-\Delta\phi^{\prime}_\mathrm{init}$ in momentum space (right axis) for the electron at the tunnel exit. The upper horizontal axis relates the tunneling angle $\gamma$ to the electron emission angle in the far-field $\beta = \gamma - 90^\circ$, neglecting Coulomb interaction after tunneling. Here, $\beta$ is defined as the angle between the final electron momentum vector and the molecular axis. (d)~Theoretical prediction for the Wigner time delay $\Delta\tau_{\mathrm{W,M}}$. The time delay is shown as a function of the electron energy $E$ and the electron's emission angle $\beta$. For the model shown (labeled `simple PFT model') the position offset at the tunnel exit $\Delta s_{\perp}$ translates directly to a time delay using the relation $\Delta\tau_{\mathrm{W,M}} = -\Delta s_{\perp} m_\mathrm{e} / \vert{}\mathbf{p}_{\mathrm{elec}}\vert{}$, which builds on the approximation that the position offset is always parallel or anti-parallel to the final electron momentum.}
\end{figure}

Figure~\ref{fig1} illustrates the expected Wigner time delays for the strong-field ionization of CO based on the partial Fourier transform (PFT) model~\cite{Murray2010,Liu_2016} and the subsequent streaking of the electronic wave packet by the driving laser field~\cite{Eckart_HASEtheo}. The underlying idea is that a position offset at the tunnel exit parallel [antiparallel] to the final electron momentum results in negative [positive] time delays. Inspection of the bound electronic wave function of CO in position space \footnote{Fig.~\ref{fig1}(a) shows a two-dimensional logarithmic density distribution $\vert{}\psi_r(x,y)\vert{}^2$ which is described by the asymptotic part of the $5\sigma$ highest occupied molecular orbital (HOMO) of CO in position space, modeled by $\psi^\mathrm{(m)}_\mathrm{n}(\mathbf{r})= \sum_{l} C_{lm} r^{(Z_{c}/\eta)-1} e^{-\eta r} Y_{lm}(\theta, \varphi)$ of the molecular Ammosov-Delone-Krainov (MO-ADK) model from Ref.~\cite{Zhao2010PRA} (with the ionization potential $I_\mathrm{p} = 14.01$\nobreak\,eV, the bond length $R=1.128\nobreak\,\text{\AA}$, $\eta=\sqrt{2I_\mathrm{p}}$, and the structure parameters $C_\mathrm{00}=2.32$, $C_\mathrm{10}=1.62$, $C_\mathrm{20}=0.82$, $C_\mathrm{30}=0.17$, and $C_\mathrm{40}=0.05$).} reveals that the electron density at the tunnel exit depends on the tunneling direction, which is specified by the angle $\gamma$ relative to the molecular axis in the plane of polarization such that $\gamma = 0^\circ$ points toward the carbon atom and $\gamma = 180^\circ$ toward the oxygen atom. Evaluating the one-dimensional (1D) projection of $\vert{}\psi_r(x,y)\vert{}^2$ perpendicular to the tunnel exit along $s_\perp$ (see Figs.~\ref{fig1}(a) and \ref{fig1}(b)) reveals position offsets $\Delta s_\perp$ that vary as a function of $\gamma$. A striking difference is that for a tunneling angle of $\gamma = 90^\circ$ we find a maximum Wigner time delay for CO compared to a Wigner time delay of $0$\nobreak\,as for all homonuclear molecules. Here, the value of the tunnel exit position of $r_0 = 9.9\nobreak\,\text{\AA}$ is calculated in parabolic coordinates~\cite{Pfeiffer2012,Hofmann2013} using the parameters of the driving laser electric field in our experiment (see below) and the field-free ionization potential $I_\mathrm{p}$. Since induced dipole and Stark shifts alter the resulting position offsets $\Delta s_\perp$ by less than $0.5\%$, we do not include them in our current work for the sake of simplicity. For tunneling along the molecular axis $\gamma = 0^\circ$ and $\gamma = 180^\circ$, the 1D projections $\vert{}\psi_r(x,y)\vert{}^2$ along $s_\perp$ are centered at $s_\perp = 0\nobreak\,\text{\AA}$ due to the cylindrical symmetry of the 5$\sigma$ HOMO around the molecular axis. In contrast, the asymmetric electron density distribution between the carbon and oxygen atoms shifts the peak position of $\vert{}\psi_r(x,y)\vert{}^2$ by up to $\pm0.8\nobreak\,\text{\AA}$ for tunnel exits that are close to the carbon atom (see e.g. $\gamma = \pm78^\circ$) and by $\mp0.2\nobreak\,\text{\AA}$ close to the oxygen atom (see e.g. $\gamma = \pm154^\circ$).
 
\begin{figure}[t]
\includegraphics[width=\columnwidth]{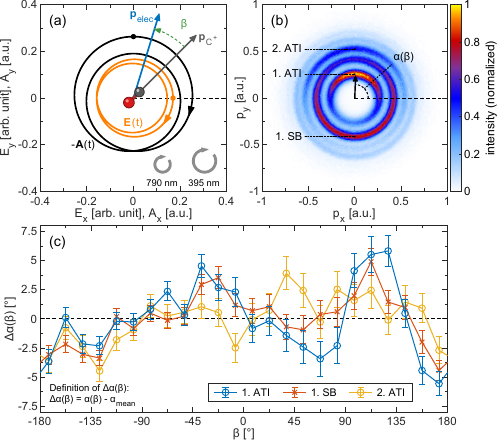}
\caption{\label{fig2} Experimental data on the holographic angular streaking of electrons (HASE) from carbon monoxide. (a) Combined laser electric field $\mathbf{E}(t)$ and corresponding negative vector potential $-\mathbf{A}(t)$. The gray arrows indicate the helicity of the low- [high-]intensity circularly polarized pulses at a central wavelength of 790\nobreak\,nm [395\nobreak\,nm]. The emission angle $\beta$ is defined as the angle from the momentum vector of the $\text{C}^+$ ion, $\mathbf{p}_{\mathrm{C^+}}$, to the final electron momentum vector $\mathbf{p}_{\mathrm{elec}}$ in the plane of polarization. (b) Measured electron momentum distribution in the laser's polarization plane, exhibiting the characteristic alternating half-ring pattern. Indicated are the first two above-threshold ionization (ATI) peaks and the first sideband (SB) peak. For each of the peaks, the most probable emission direction $\alpha(\beta)$ is determined. (c) Variations of the most probable emission direction as a function of $\beta$, $\Delta\alpha(\beta)$ (see text). The error bars show the statistical errors only.}
\end{figure}

By means of the Fourier transform a position offset $\Delta s_\perp$ along the direction $s_\perp$ translates into a linear phase gradient in momentum space. This leads  to the relation $\Delta\phi^{\prime}_\mathrm{init} = -\Delta s_\perp/\hbar$, as visualized in Fig.~\ref{fig1}(c). We define the emission angle $\beta$ as the angle between the final electron momentum vector $\mathbf{p}_{\mathrm{elec}}$ and the momentum vector of the $\mathrm{C}^+$ ion in the plane of polarization (see Fig.~\ref{fig2}(a)). Neglecting Coulomb interaction in the dynamics after tunneling, $\mathbf{p}_\mathrm{elec}$ is perpendicular to the tunneling direction. Therefore, $\beta$ in momentum space is related to the tunneling angle $\gamma$ in position space according to $\beta = \gamma - 90^\circ$ for the given helicity of the circularly polarized driving field that is used in our experiment. Consequently, the position offset $\Delta s_\perp$ directly alters the path length of the electron from the tunnel exit position to the detector \cite{Eckart_HASEtheo} and thereby translates directly into a time delay $\Delta\tau_{\mathrm{W,M}} = -\frac{ m_\mathrm{e}} {|\mathbf{p}_{\mathrm{elec}}|} \Delta s_{\perp} = \frac{\hbar m_\mathrm{e}}{|\mathbf{p}_\mathrm{elec}|}\Delta\phi^{\prime}_\mathrm{init}$. Based on this simple model, the values for $\Delta\tau_{\mathrm{W,M}}$ are shown in Fig.~\ref{fig1}(d), which predicts values up to $\pm136$\nobreak\,as for a tunnel exit close to the carbon atom (see $\beta = -168^\circ$ or $\beta =-12^\circ$) and $\mp31$\nobreak\,as for tunnel exits close to the oxygen atom (see $\beta = +116^\circ$ or $\beta = +64^\circ$). 
 
Experimentally, we investigate the single ionization and subsequent dissociation of CO using a co-rotating two-color (CoRTC) laser field and a cold target recoil ion momentum spectroscopy (COLTRIMS) reaction microscope~\cite{Dorner2000,Ullrich_2003}. To generate the CoRTC field required for the experiment, pulses from a Ti:sapphire femtosecond laser system (100\nobreak\,kHz, Wyvern-500, KMLabs) operating at a central wavelength of 790\nobreak\,nm and a pulse duration of 50\nobreak\,fs are split to propagate along two pathways in an interferometric setup. In one pathway, a 200\nobreak\,µm $\beta$-barium borate crystal is used to frequency-double the laser pulses. The intensity and the polarization of both single-color laser pulses can be adjusted independently. The relative phase between the two single-color laser pulses is controlled by a piezo delay stage. After a beam combiner, the two-color laser beam is focused by a spherical mirror (focal length $f = 60$\nobreak\,mm) into a cold molecular beam of CO. The electron and carbon ion ($\mathrm{C}^+$) released in the ionization and subsequent dissociation of a CO molecule are guided by a homogeneous electric field of 26.8\nobreak\,V/cm and a homogeneous magnetic field of 9.4\nobreak\,Gauss onto position- and time-sensitive detectors~\cite{Jagutzki_2002}. The lengths of the spectrometer are 30.6\nobreak\,cm and 17.0\nobreak\,cm for the electron and ion arms, respectively. The detectors consist of microchannel plates with a diameter of 80\nobreak\,mm and a hexagonal delay-line anode. The time-of-flight and the position-of-impact of the electron and the ion are measured in coincidence, which allows for obtaining three-dimensional momentum distributions of all charged particles~\cite{Dorner2000,Ullrich_2003}.

Figure~\ref{fig2}(a) shows the Lissajous curve of the combined two-color laser field and the corresponding negative vector potential. The laser pulse at a central wavelength of 790\nobreak\,nm [395\nobreak\,nm] has an intensity of $3.1\times10^{11}$\nobreak\,W/cm$^2$ [$4.8\times10^{13}$\nobreak\,W/cm$^2$], which corresponds to a peak electric field of $E_\mathrm{790}=0.0021$\nobreak\,a.u. [$E_\mathrm{395}=0.0262$\nobreak\,a.u.]. For the peak electric field of the emerging CoRTC field this leads to a Keldysh parameter of $\gamma_\mathrm{395}= \frac{\omega_\mathrm{395}\sqrt{2I_\mathrm{p}}}{E_\mathrm{395}} = 4.5$, using the ionization potential $I_\mathrm{p}=14.01\nobreak\,\mathrm{eV}$ and the light's angular frequency of $\omega_\mathrm{395}=0.115$\nobreak\,a.u. Both single-color intensities are calibrated separately using circularly polarized light. The intensity of the 790\nobreak\,nm laser pulse was determined by comparing the measured distribution of the radial electron momentum $p_r$ with a semiclassical two-step (SCTS) simulation (as in Ref.~\cite{Eckart2018_Offsets}). For the 395\nobreak\,nm contribution, the intensity was calibrated as in Ref.~\cite{Eckart2016}. Due to the very different intensities of the single-color laser pulses, the component of the driving field at 395\nobreak\,nm governs the ionization dynamics while the component at 790\nobreak\,nm slightly perturbs this process. 

Figure~\ref{fig2}(b) shows the measured electron momentum distribution for the dissociating channel of $\mathrm{CO} \rightarrow \mathrm{C}^+ + \mathrm{O} + e^-$ in the plane of polarization selected within a kinetic energy release (KER) range of $0.92\nobreak\,\mathrm{eV} \leq \mathrm{KER} \leq 1.07\nobreak\,\mathrm{eV}$ and restricted to molecular orientations within $\pm 30^\circ$ relative to the polarization plane.  The distribution is integrated over all $\beta$. As expected, the alternating half-ring structure of HASE experiments is visible, where the above-threshold ionization (ATI) peaks are the peaks that would remain in the absence of the 790\nobreak\,nm field. The other electron energy peaks are referred to as sideband (SB) peaks. To analyze the interference pattern, the envelope is extracted by low-pass filtering of an overall measured energy distribution (as in Ref.~\cite{Trabert2021}). The interference pattern is obtained by bin-wise division of the measured momentum distribution by the extracted intensity envelope. This decomposition is performed for each value of $\beta$ individually. We determine the most probable emission direction $\alpha(\beta)$ for the first two ATI peaks and the first SB peak from our normalized momentum distribution for every $\beta$. The variation of $\alpha(\beta)$ is defined as $\Delta\alpha(\beta) = \alpha(\beta) - \alpha_{\mathrm{mean}}$, where $\alpha_{\mathrm{mean}}$ is the mean of $\alpha(\beta)$ for all values of $\beta$ for a given electron energy peak. The results for $\Delta\alpha(\beta)$ are depicted in Fig.~\ref{fig2}(c). This procedure is analogous to that described in Refs.~\cite{Trabert2021,Trabert2023}. Statistical uncertainties of the extracted most probable emission angles $\alpha(\beta)$ were estimated using a parametric bootstrap procedure with 1000 Poisson-resampled datasets. The statistical error shown in Fig. \ref{fig2}(c) is the standard deviation of the extracted angles $\alpha(\beta)$ across the resampled datasets.

The measured values of $\Delta\alpha(\beta)$ displayed in Fig.~\ref{fig2}(c) can be linked to momentum space phase gradients $\Delta\phi^{\prime}_{\mathrm{init}}$ by means of an SCTS simulation. In the SCTS, the post-tunneling dynamics of the electron are simulated by computing classical electron trajectories that start at the tunnel exit calculated in parabolic coordinates~\cite{Pfeiffer2012,Hofmann2013} (neglecting induced dipole and Stark shifts as above). The electrons are subsequently driven by the laser electric fields in a $-1/r$ potential, where $r$ is the distance of the electron to the center of mass of the CO$^+$ ion. The SCTS simulation can reproduce the alternating half-ring pattern because, for each electron trajectory, the semiclassical phase is also evaluated. We calculated $250 \times 10^6$ trajectories, with a two-cycle flat-top CoRTC pulse with the same time-dependent electric fields as in Fig.~\ref{fig2}(a). In the simulation, the implementation of a phase $\phi_{\mathrm{init}} = p_{\mathrm{init}}\Delta\phi^{\prime}_{\mathrm{init}}$ that depends linearly on the initial momentum $p_{\mathrm{init}}$ perpendicular to the tunneling direction causes the alternating half-ring pattern to rotate by $\Delta\alpha$ as in Ref. \cite{Eckart_HASEtheo}. Thus, a phase gradient $\Delta\phi^{\prime}_{\mathrm{init}}$ can be assigned to each measured value of $\Delta\alpha(\beta)$ for the respective energy peak. Furthermore, after tunneling, the attractive Coulomb interaction between the electron and the remaining ion leads to an additional rotation by an angle $\kappa$ in the polarization plane~\cite{Eckle2008}. To relate the measured emission angle $\beta$ in momentum space to the tunneling angle $\gamma$ in position space, we use the classical mapping $\gamma = \beta + 90^\circ - \kappa$. Here, the deflection angles of $\kappa_{\mathrm{1.ATI}} = -24.6^\circ$, $\kappa_{\mathrm{1.SB}} = -14.0^\circ$, and $\kappa_{\mathrm{2.ATI}} = -9.9^\circ$ for the three respective electron energy peaks are extracted from our SCTS simulation as in Ref.~\cite{Trabert2021}.

\begin{figure}[t]
\includegraphics[width=\columnwidth]{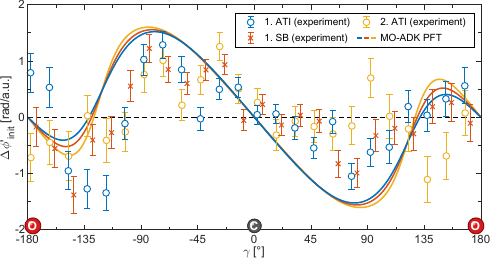}
\caption{\label{fig3} Comparison of the experimentally obtained phase gradient $\Delta\phi^{\prime}_{\mathrm{init}}$ and the corresponding predictions of the MO-ADK PFT model (solid lines) at the tunnel exit. Error bars indicate statistical uncertainties only.}
\end{figure}

The experimentally obtained phase gradients $\Delta\phi^{\prime}_{\mathrm{init}}$ for three electron energy peaks are shown in Fig.~\ref{fig3} as a function of $\gamma$. The results are compared to the extracted phase gradients from the MO-ADK PFT model (solid lines), which takes the electronic wave packet at the tunnel exit from the molecular Ammosov-Delone-Krainov (MO-ADK) model in Ref.~\cite{Zhao2010PRA} as an input for the PFT approach \cite{Murray2010,Liu_2016}. Within this model, the momentum-space phase gradients of the electronic wave function at the tunnel exit, evaluated at the corresponding initial momentum $p_\mathrm{init}$, are obtained by calculating the Fourier transform of the 1D spatial slices of $\Psi_r(s_{\perp})$ perpendicular to the tunnel direction, as visualized in Fig.~\ref{fig1}(a) for the `simple PFT model'. The experimental data for the first ATI and SB peaks exhibit an asymmetric behavior with respect to the molecular axis orientation, with phase gradients of up to $\pm1.5$\nobreak\,rad/a.u. close to the carbon side ($\gamma \approx \mp 78^\circ$) and smaller amplitudes of about $\pm0.5$\nobreak\,rad/a.u. near the oxygen side ($\gamma \approx \pm 154^\circ$). This asymmetry is well reproduced within the MO-ADK PFT model. 

The initial phase gradients can be used to obtain the Wigner time delay $\Delta\tau_{\mathrm{W,M}}$ in the continuum Ref.~\cite{Eckart_HASEtheo}:
\begin{equation}
\Delta \tau_\mathrm{W,M} (E,\beta) \approx \frac{\hbar m_\mathrm{e}}{p_\mathrm{elec}} \Delta \phi^{\prime}_\mathrm{init}(p_\mathrm{init}(p_\mathrm{elec}),\gamma(\beta)) \frac{dp_\mathrm{init}}{dp_\mathrm{elec}}
\end{equation}
The mapping functions $p_\mathrm{init}(p_\mathrm{elec})$~\footnote{The simulated momentum mapping can be parameterized as $p_\mathrm{init}(p_\mathrm{elec}) = \sqrt{p_\mathrm{elec}^2 + b^2} - c$ with $b = 0.36\,\mathrm{a.u.}$ and $c = 0.275\,\mathrm{a.u.}$}, $\gamma(\beta)$ and $\frac{dp_\mathrm{init}}{dp_\mathrm{elec}}$ are obtained from our semiclassical simulation as in Ref.~\cite{Trabert2021}.

\begin{figure}[t]
\includegraphics[width=\columnwidth]{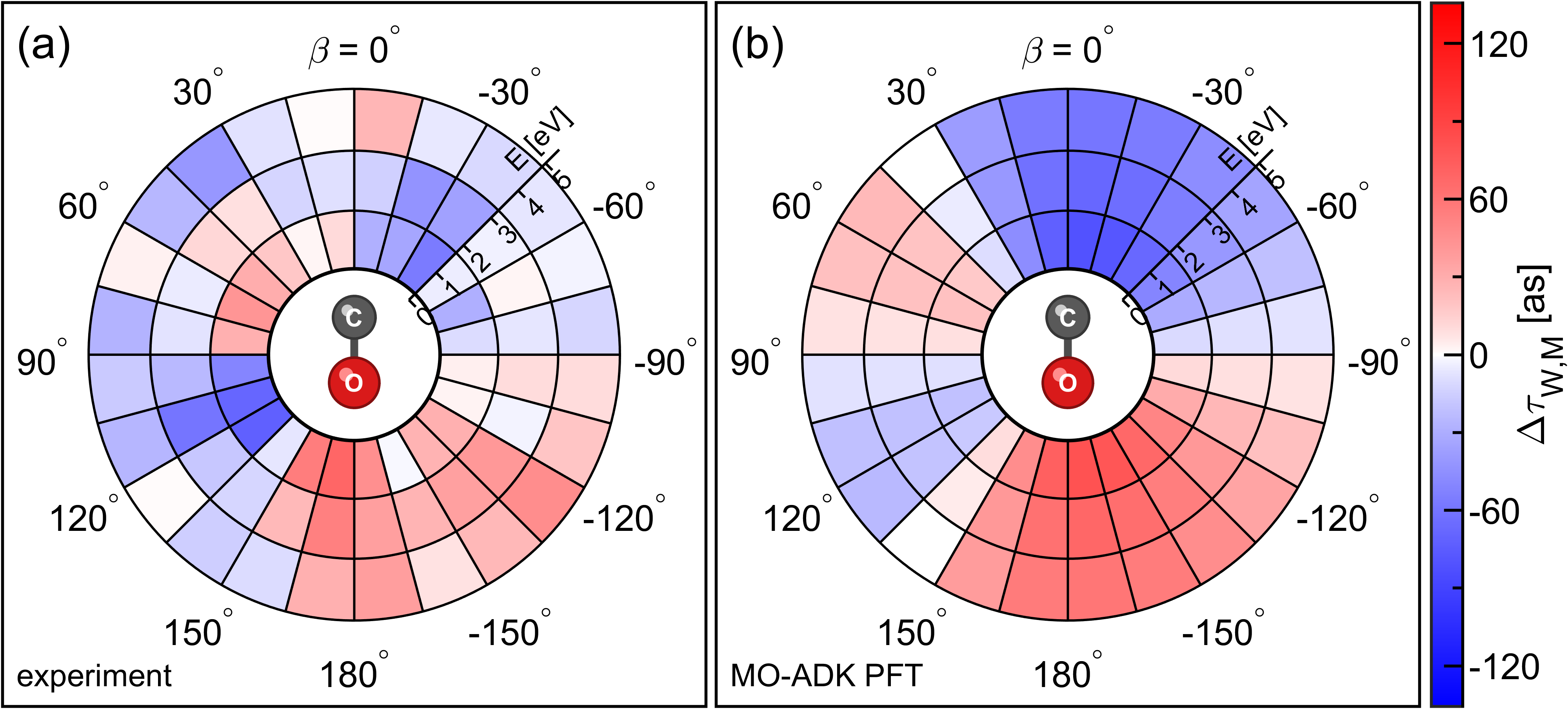}
\caption{\label{fig4} Measured and calculated changes of the Wigner time delay $\Delta\tau_{\mathrm{W,M}}$ as a function of $\beta$ in the molecular frame of carbon monoxide. (a) Experimentally retrieved Wigner time delay. (b) Corresponding Wigner time delay predicted by the MO-ADK PFT model.}
\end{figure}

Figure~\ref{fig4} compares the experimentally retrieved Wigner time delays (Fig.~\ref{fig4}(a)) to the corresponding predictions of the MO-ADK PFT model (Fig.~\ref{fig4}(b)). For the lower-energy peaks ($E_{\mathrm{1.ATI}} = 0.97$\nobreak\,eV, $E_{\mathrm{1.SB}} = 2.48$\nobreak\,eV), the experimental data in Fig.~\ref{fig4}(a) show the same angular asymmetry as the MO-ADK PFT model in Fig.~\ref{fig4}(b). The results are also in line with previous theoretic predictions by Guo \textit{et al.}~\cite{Guo2022}. In particular, the differences for electrons originating from the carbon side compared to the oxygen side are clearly resolved, demonstrating the spatial asymmetry of the $5\sigma$ HOMO of CO as expected for a stereo Wigner time delay \cite{vos2018orientation}. Overall, both the angular dependence and the magnitude of the delays comparing experiment and theory are in good quantitative agreement, yielding time delays of up to about $\pm70$\nobreak\,as.

The experimentally obtained data in the range $+30^\circ \leq \beta \leq +90^\circ$, the second ATI peak ($E_{\mathrm{2.ATI}} = 4.07$\nobreak\,eV) shows negative time delays, which disagrees with the result from the MO-ADK PFT model. Time delays at these energies were not evaluated in previous theoretical simulations by Guo \textit{et al.}~\cite{Guo2022}, which relied on a simplified two-$s$-orbital model and did not extract $\Delta\tau_\mathrm{W,M}$ directly from experimental data. There are several physical mechanisms that could lead to this discrepancy. First, the ionization rate of the $5\sigma$ HOMO exhibits a strong angular dependence. The ionization yield on the oxygen side is strongly suppressed, enhancing the relative contribution of lower-lying orbitals such as the $1\pi$ HOMO-1~\cite{Zhao2010PRA,Wu2012}, leading to contributions with a different phase structure. Second, with a Keldysh parameter of $\gamma_\mathrm{395} = 4.5$, the experiment operates well within the regime of non-adiabatic tunneling~\cite{Misha2005,Barth2011,Eckart2018_Offsets}, which can give rise to complicated dynamics in the tunnel and affect the electron's phase at the tunnel exit. Third, carbon monoxide has an anisotropic long-range ionic potential and a strong permanent dipole moment, which may induce orientation- and energy-dependent phase shifts in the continuum.

Our results can be compared to Wigner time delays of CO in the single-photon regime ~\cite{Rist2021,vos2018orientation}. Using RABBITT, Vos et al. measured stereo Wigner time delays, which relate electron emission from the carbon or oxygen sides of CO~\cite{vos2018orientation}. They also establish a connection between their measured Wigner time delays to theoretically predicted mean electron positions along the molecular axis. Using a very different approach, Rist et al. extracted angle-dependent Wigner time delays for C(1s) ionization of CO, revealing the relevance of electron scattering in the anisotropic molecular potential and shape resonances ~\cite{Rist2021}. Both Vos et al.\ and Rist et al.\ report on Wigner time delays from CO in the single-photon regime that vary on the order of $100$\nobreak\,attoseconds as a function of the electron's emission direction, which is surprisingly similar to our findings in the strong-field regime.

In conclusion, we have measured the angular dependence of the phase gradients $\Delta\phi^{\prime}_{\mathrm{init}}$ and the corresponding Wigner time delays $\Delta\tau_{\mathrm{W,M}}$ for CO upon strong-field ionization. Our findings show that the asymmetric structure of the $5\sigma$ HOMO induces strongly direction-dependent position offsets at the tunnel exit, which translate into Wigner time delays of up to $\pm70$\nobreak\,as. For the first two energy peaks, these experimental results are in good agreement with the predictions of the MO-ADK PFT model, as well as with previous simulations by Guo \textit{et al.}~\cite{Guo2022}. The experimental data show a sign reversal at higher electron energies for electrons tunneling near the oxygen side, which is not predicted by the MO-ADK PFT model. We speculate that this unexpected discrepancy observed at higher electron energies can arise from ionization contributions of lower-lying electronic states, (non-adiabatic) sub-barrier dynamics or long-range Coulomb interaction and warrants further research. Our findings show that the HASE technique enables the investigation of asymmetric orbital structures, underscoring the need for advanced theoretical models that fully capture multi-orbital, non-adiabatic tunneling and subsequent continuum dynamics for the strong-field ionization of heteronuclear molecules~\cite{Haertelt2026}.

\section{Acknowledgments}
\normalsize
The experimental work was supported by the Deutsche Forschungsgemeinschaft (DFG) - Project No. 328961117 - SFB 1319 ELCH. We thank Reinhard Dörner for fruitful discussions.

\bibliography{literatur}

\begin{thebibliography}{33}%
\makeatletter
\providecommand \@ifxundefined [1]{%
 \@ifx{#1\undefined}
}%
\providecommand \@ifnum [1]{%
 \ifnum #1\expandafter \@firstoftwo
 \else \expandafter \@secondoftwo
 \fi
}%
\providecommand \@ifx [1]{%
 \ifx #1\expandafter \@firstoftwo
 \else \expandafter \@secondoftwo
 \fi
}%
\providecommand \natexlab [1]{#1}%
\providecommand \enquote  [1]{``#1''}%
\providecommand \bibnamefont  [1]{#1}%
\providecommand \bibfnamefont [1]{#1}%
\providecommand \citenamefont [1]{#1}%
\providecommand \href@noop [0]{\@secondoftwo}%
\providecommand \href [0]{\begingroup \@sanitize@url \@href}%
\providecommand \@href[1]{\@@startlink{#1}\@@href}%
\providecommand \@@href[1]{\endgroup#1\@@endlink}%
\providecommand \@sanitize@url [0]{\catcode `\\12\catcode `\$12\catcode
  `\&12\catcode `\#12\catcode `\^12\catcode `\_12\catcode `\%12\relax}%
\providecommand \@@startlink[1]{}%
\providecommand \@@endlink[0]{}%
\providecommand \url  [0]{\begingroup\@sanitize@url \@url }%
\providecommand \@url [1]{\endgroup\@href {#1}{\urlprefix }}%
\providecommand \urlprefix  [0]{URL }%
\providecommand \Eprint [0]{\href }%
\providecommand \doibase [0]{https://doi.org/}%
\providecommand \selectlanguage [0]{\@gobble}%
\providecommand \bibinfo  [0]{\@secondoftwo}%
\providecommand \bibfield  [0]{\@secondoftwo}%
\providecommand \translation [1]{[#1]}%
\providecommand \BibitemOpen [0]{}%
\providecommand \bibitemStop [0]{}%
\providecommand \bibitemNoStop [0]{.\EOS\space}%
\providecommand \EOS [0]{\spacefactor3000\relax}%
\providecommand \BibitemShut  [1]{\csname bibitem#1\endcsname}%
\let\auto@bib@innerbib\@empty
\bibitem [{\citenamefont {Wigner}(1955)}]{Wigner1954}%
  \BibitemOpen
  \bibfield  {author} {\bibinfo {author} {\bibfnamefont {E.~P.}\ \bibnamefont
  {Wigner}},\ }\bibfield  {title} {\bibinfo {title} {{Lower Limit for the
  Energy Derivative of the Scattering Phase Shift}},\ }\href
  {https://doi.org/10.1103/PhysRev.98.145} {\bibfield  {journal} {\bibinfo
  {journal} {Phys. Rev.}\ }\textbf {\bibinfo {volume} {98}},\ \bibinfo {pages}
  {145} (\bibinfo {year} {1955})}\BibitemShut {NoStop}%
\bibitem [{\citenamefont {Schultze}\ \emph {et~al.}(2010)\citenamefont
  {Schultze}, \citenamefont {Fie{\ss}}, \citenamefont {Karpowicz},
  \citenamefont {Gagnon}, \citenamefont {Korbman}, \citenamefont {Hofstetter},
  \citenamefont {Neppl}, \citenamefont {Cavalieri}, \citenamefont {Komninos},
  \citenamefont {Mercouris}, \citenamefont {Nicolaides}, \citenamefont
  {Pazourek}, \citenamefont {Nagele}, \citenamefont {Feist}, \citenamefont
  {Burgd{\"{o}}rfer}, \citenamefont {Azzeer}, \citenamefont {Ernstorfer},
  \citenamefont {Kienberger}, \citenamefont {Kleineberg}, \citenamefont
  {Goulielmakis}, \citenamefont {Krausz},\ and\ \citenamefont
  {Yakovlev}}]{Schultze2010}%
  \BibitemOpen
  \bibfield  {author} {\bibinfo {author} {\bibfnamefont {M.}~\bibnamefont
  {Schultze}}, \bibinfo {author} {\bibfnamefont {M.}~\bibnamefont {Fie{\ss}}},
  \bibinfo {author} {\bibfnamefont {N.}~\bibnamefont {Karpowicz}}, \bibinfo
  {author} {\bibfnamefont {J.}~\bibnamefont {Gagnon}}, \bibinfo {author}
  {\bibfnamefont {M.}~\bibnamefont {Korbman}}, \bibinfo {author} {\bibfnamefont
  {M.}~\bibnamefont {Hofstetter}}, \bibinfo {author} {\bibfnamefont
  {S.}~\bibnamefont {Neppl}}, \bibinfo {author} {\bibfnamefont {A.~L.}\
  \bibnamefont {Cavalieri}}, \bibinfo {author} {\bibfnamefont {Y.}~\bibnamefont
  {Komninos}}, \bibinfo {author} {\bibfnamefont {T.}~\bibnamefont {Mercouris}},
  \bibinfo {author} {\bibfnamefont {C.~A.}\ \bibnamefont {Nicolaides}},
  \bibinfo {author} {\bibfnamefont {R.}~\bibnamefont {Pazourek}}, \bibinfo
  {author} {\bibfnamefont {S.}~\bibnamefont {Nagele}}, \bibinfo {author}
  {\bibfnamefont {J.}~\bibnamefont {Feist}}, \bibinfo {author} {\bibfnamefont
  {J.}~\bibnamefont {Burgd{\"{o}}rfer}}, \bibinfo {author} {\bibfnamefont
  {A.~M.}\ \bibnamefont {Azzeer}}, \bibinfo {author} {\bibfnamefont
  {R.}~\bibnamefont {Ernstorfer}}, \bibinfo {author} {\bibfnamefont
  {R.}~\bibnamefont {Kienberger}}, \bibinfo {author} {\bibfnamefont
  {U.}~\bibnamefont {Kleineberg}}, \bibinfo {author} {\bibfnamefont
  {E.}~\bibnamefont {Goulielmakis}}, \bibinfo {author} {\bibfnamefont
  {F.}~\bibnamefont {Krausz}},\ and\ \bibinfo {author} {\bibfnamefont {V.~S.}\
  \bibnamefont {Yakovlev}},\ }\bibfield  {title} {\bibinfo {title} {{Delay in
  Photoemission}},\ }\href {https://doi.org/10.1126/science.1189401} {\bibfield
   {journal} {\bibinfo  {journal} {Science}\ }\textbf {\bibinfo {volume}
  {328}},\ \bibinfo {pages} {1658} (\bibinfo {year} {2010})}\BibitemShut
  {NoStop}%
\bibitem [{\citenamefont {Kl{\"{u}}nder}\ \emph {et~al.}(2011)\citenamefont
  {Kl{\"{u}}nder}, \citenamefont {Dahlstr{\"{o}}m}, \citenamefont
  {Gisselbrecht}, \citenamefont {Fordell}, \citenamefont {Swoboda},
  \citenamefont {Gu{\'{e}}not}, \citenamefont {Johnsson}, \citenamefont
  {Caillat}, \citenamefont {Mauritsson}, \citenamefont {Maquet}, \citenamefont
  {Ta{\"{i}}eb},\ and\ \citenamefont {L'Huillier}}]{Klunder2011}%
  \BibitemOpen
  \bibfield  {author} {\bibinfo {author} {\bibfnamefont {K.}~\bibnamefont
  {Kl{\"{u}}nder}}, \bibinfo {author} {\bibfnamefont {J.~M.}\ \bibnamefont
  {Dahlstr{\"{o}}m}}, \bibinfo {author} {\bibfnamefont {M.}~\bibnamefont
  {Gisselbrecht}}, \bibinfo {author} {\bibfnamefont {T.}~\bibnamefont
  {Fordell}}, \bibinfo {author} {\bibfnamefont {M.}~\bibnamefont {Swoboda}},
  \bibinfo {author} {\bibfnamefont {D.}~\bibnamefont {Gu{\'{e}}not}}, \bibinfo
  {author} {\bibfnamefont {P.}~\bibnamefont {Johnsson}}, \bibinfo {author}
  {\bibfnamefont {J.}~\bibnamefont {Caillat}}, \bibinfo {author} {\bibfnamefont
  {J.}~\bibnamefont {Mauritsson}}, \bibinfo {author} {\bibfnamefont
  {A.}~\bibnamefont {Maquet}}, \bibinfo {author} {\bibfnamefont
  {R.}~\bibnamefont {Ta{\"{i}}eb}},\ and\ \bibinfo {author} {\bibfnamefont
  {A.}~\bibnamefont {L'Huillier}},\ }\bibfield  {title} {\bibinfo {title}
  {{Probing Single-Photon Ionization on the Attosecond Time Scale}},\ }\href
  {https://doi.org/10.1103/PhysRevLett.106.143002} {\bibfield  {journal}
  {\bibinfo  {journal} {Phys. Rev. Lett.}\ }\textbf {\bibinfo {volume} {106}},\
  \bibinfo {pages} {143002} (\bibinfo {year} {2011})}\BibitemShut {NoStop}%
\bibitem [{\citenamefont {Gruson}\ \emph {et~al.}(2016)\citenamefont {Gruson},
  \citenamefont {Barreau}, \citenamefont {Jim{\'e}nez-Gal{\'a}n}, \citenamefont
  {Risoud}, \citenamefont {Caillat}, \citenamefont {Maquet}, \citenamefont
  {Carr{\'e}}, \citenamefont {Lepetit}, \citenamefont {Hergott}, \citenamefont
  {Ruchon}, \citenamefont {Argenti}, \citenamefont {Ta{\"i}eb}, \citenamefont
  {Mart{\'i}n},\ and\ \citenamefont {Sali{\`e}res}}]{Gruson2016}%
  \BibitemOpen
  \bibfield  {author} {\bibinfo {author} {\bibfnamefont {V.}~\bibnamefont
  {Gruson}}, \bibinfo {author} {\bibfnamefont {L.}~\bibnamefont {Barreau}},
  \bibinfo {author} {\bibfnamefont {{\'A}.}~\bibnamefont
  {Jim{\'e}nez-Gal{\'a}n}}, \bibinfo {author} {\bibfnamefont {F.}~\bibnamefont
  {Risoud}}, \bibinfo {author} {\bibfnamefont {J.}~\bibnamefont {Caillat}},
  \bibinfo {author} {\bibfnamefont {A.}~\bibnamefont {Maquet}}, \bibinfo
  {author} {\bibfnamefont {B.}~\bibnamefont {Carr{\'e}}}, \bibinfo {author}
  {\bibfnamefont {F.}~\bibnamefont {Lepetit}}, \bibinfo {author} {\bibfnamefont
  {J.-F.}\ \bibnamefont {Hergott}}, \bibinfo {author} {\bibfnamefont
  {T.}~\bibnamefont {Ruchon}}, \bibinfo {author} {\bibfnamefont
  {L.}~\bibnamefont {Argenti}}, \bibinfo {author} {\bibfnamefont
  {R.}~\bibnamefont {Ta{\"i}eb}}, \bibinfo {author} {\bibfnamefont
  {F.}~\bibnamefont {Mart{\'i}n}},\ and\ \bibinfo {author} {\bibfnamefont
  {P.}~\bibnamefont {Sali{\`e}res}},\ }\bibfield  {title} {\bibinfo {title}
  {Attosecond dynamics through a {Fano} resonance: Monitoring the birth of a
  photoelectron},\ }\href {https://doi.org/10.1126/science.aah5188} {\bibfield
  {journal} {\bibinfo  {journal} {Science}\ }\textbf {\bibinfo {volume}
  {354}},\ \bibinfo {pages} {734} (\bibinfo {year} {2016})}\BibitemShut
  {NoStop}%
\bibitem [{\citenamefont {Heuser}\ \emph {et~al.}(2016)\citenamefont {Heuser},
  \citenamefont {Jim{\'e}nez~Gal{\'a}n}, \citenamefont {Cirelli}, \citenamefont
  {Marante}, \citenamefont {Sabbar}, \citenamefont {Boge}, \citenamefont
  {Lucchini}, \citenamefont {Gallmann}, \citenamefont {Ivanov}, \citenamefont
  {Kheifets}, \citenamefont {Dahlstr{\"o}m}, \citenamefont {Lindroth},
  \citenamefont {Argenti}, \citenamefont {Mart{\'{\i}}n},\ and\ \citenamefont
  {Keller}}]{Heuser2016}%
  \BibitemOpen
  \bibfield  {author} {\bibinfo {author} {\bibfnamefont {S.}~\bibnamefont
  {Heuser}}, \bibinfo {author} {\bibfnamefont {{\'A}.}~\bibnamefont
  {Jim{\'e}nez~Gal{\'a}n}}, \bibinfo {author} {\bibfnamefont {C.}~\bibnamefont
  {Cirelli}}, \bibinfo {author} {\bibfnamefont {C.}~\bibnamefont {Marante}},
  \bibinfo {author} {\bibfnamefont {M.}~\bibnamefont {Sabbar}}, \bibinfo
  {author} {\bibfnamefont {R.}~\bibnamefont {Boge}}, \bibinfo {author}
  {\bibfnamefont {M.}~\bibnamefont {Lucchini}}, \bibinfo {author}
  {\bibfnamefont {L.}~\bibnamefont {Gallmann}}, \bibinfo {author}
  {\bibfnamefont {I.}~\bibnamefont {Ivanov}}, \bibinfo {author} {\bibfnamefont
  {A.~S.}\ \bibnamefont {Kheifets}}, \bibinfo {author} {\bibfnamefont {J.~M.}\
  \bibnamefont {Dahlstr{\"o}m}}, \bibinfo {author} {\bibfnamefont
  {E.}~\bibnamefont {Lindroth}}, \bibinfo {author} {\bibfnamefont
  {L.}~\bibnamefont {Argenti}}, \bibinfo {author} {\bibfnamefont
  {F.}~\bibnamefont {Mart{\'{\i}}n}},\ and\ \bibinfo {author} {\bibfnamefont
  {U.}~\bibnamefont {Keller}},\ }\bibfield  {title} {\bibinfo {title} {Angular
  dependence of photoemission time delay in helium},\ }\href
  {https://doi.org/10.1103/PhysRevA.94.063409} {\bibfield  {journal} {\bibinfo
  {journal} {Phys. Rev. A}\ }\textbf {\bibinfo {volume} {94}},\ \bibinfo
  {pages} {063409} (\bibinfo {year} {2016})}\BibitemShut {NoStop}%
\bibitem [{\citenamefont {Vos}\ \emph {et~al.}(2018)\citenamefont {Vos},
  \citenamefont {Cattaneo}, \citenamefont {Patchkovskii}, \citenamefont
  {Zimmermann}, \citenamefont {Cirelli}, \citenamefont {Lucchini},
  \citenamefont {Kheifets}, \citenamefont {Landsman},\ and\ \citenamefont
  {Keller}}]{vos2018orientation}%
  \BibitemOpen
  \bibfield  {author} {\bibinfo {author} {\bibfnamefont {J.}~\bibnamefont
  {Vos}}, \bibinfo {author} {\bibfnamefont {L.}~\bibnamefont {Cattaneo}},
  \bibinfo {author} {\bibfnamefont {S.}~\bibnamefont {Patchkovskii}}, \bibinfo
  {author} {\bibfnamefont {T.}~\bibnamefont {Zimmermann}}, \bibinfo {author}
  {\bibfnamefont {C.}~\bibnamefont {Cirelli}}, \bibinfo {author} {\bibfnamefont
  {M.}~\bibnamefont {Lucchini}}, \bibinfo {author} {\bibfnamefont
  {A.}~\bibnamefont {Kheifets}}, \bibinfo {author} {\bibfnamefont {A.~S.}\
  \bibnamefont {Landsman}},\ and\ \bibinfo {author} {\bibfnamefont
  {U.}~\bibnamefont {Keller}},\ }\bibfield  {title} {\bibinfo {title}
  {{Orientation-dependent stereo Wigner time delay and electron localization in
  a small molecule}},\ }\href {https://doi.org/10.1126/science.aao4731}
  {\bibfield  {journal} {\bibinfo  {journal} {Science}\ }\textbf {\bibinfo
  {volume} {360}},\ \bibinfo {pages} {1326} (\bibinfo {year}
  {2018})}\BibitemShut {NoStop}%
\bibitem [{\citenamefont {Gong}\ \emph {et~al.}(2022)\citenamefont {Gong},
  \citenamefont {Jiang}, \citenamefont {Tong}, \citenamefont {Qiang},
  \citenamefont {Lu}, \citenamefont {Ni}, \citenamefont {Lucchese},
  \citenamefont {Ueda},\ and\ \citenamefont {Wu}}]{Gong2022}%
  \BibitemOpen
  \bibfield  {author} {\bibinfo {author} {\bibfnamefont {X.}~\bibnamefont
  {Gong}}, \bibinfo {author} {\bibfnamefont {W.}~\bibnamefont {Jiang}},
  \bibinfo {author} {\bibfnamefont {J.}~\bibnamefont {Tong}}, \bibinfo {author}
  {\bibfnamefont {J.}~\bibnamefont {Qiang}}, \bibinfo {author} {\bibfnamefont
  {P.}~\bibnamefont {Lu}}, \bibinfo {author} {\bibfnamefont {H.}~\bibnamefont
  {Ni}}, \bibinfo {author} {\bibfnamefont {R.}~\bibnamefont {Lucchese}},
  \bibinfo {author} {\bibfnamefont {K.}~\bibnamefont {Ueda}},\ and\ \bibinfo
  {author} {\bibfnamefont {J.}~\bibnamefont {Wu}},\ }\bibfield  {title}
  {\bibinfo {title} {{Asymmetric Attosecond Photoionization in Molecular Shape
  Resonance}},\ }\href {https://doi.org/10.1103/PhysRevX.12.011002} {\bibfield
  {journal} {\bibinfo  {journal} {Phys. Rev. X}\ }\textbf {\bibinfo {volume}
  {12}},\ \bibinfo {pages} {011002} (\bibinfo {year} {2022})}\BibitemShut
  {NoStop}%
\bibitem [{\citenamefont {Rist}\ \emph {et~al.}(2021)\citenamefont {Rist},
  \citenamefont {Klyssek}, \citenamefont {Novikovskiy}, \citenamefont
  {Kircher}, \citenamefont {Vela-P{\'{e}}rez}, \citenamefont {Trabert},
  \citenamefont {Grundmann}, \citenamefont {Tsitsonis}, \citenamefont
  {Siebert}, \citenamefont {Geyer}, \citenamefont {Melzer}, \citenamefont
  {Schwarz}, \citenamefont {Anders}, \citenamefont {Kaiser}, \citenamefont
  {Fehre}, \citenamefont {Hartung}, \citenamefont {Eckart}, \citenamefont {{{L.
  Ph. H. Schmidt}}}, \citenamefont {Sch{\"{o}}ffler}, \citenamefont {Davis},
  \citenamefont {Williams}, \citenamefont {Trinter}, \citenamefont
  {D{\"{o}}rner}, \citenamefont {Demekhin},\ and\ \citenamefont
  {Jahnke}}]{Rist2021}%
  \BibitemOpen
  \bibfield  {author} {\bibinfo {author} {\bibfnamefont {J.}~\bibnamefont
  {Rist}}, \bibinfo {author} {\bibfnamefont {K.}~\bibnamefont {Klyssek}},
  \bibinfo {author} {\bibfnamefont {N.~M.}\ \bibnamefont {Novikovskiy}},
  \bibinfo {author} {\bibfnamefont {M.}~\bibnamefont {Kircher}}, \bibinfo
  {author} {\bibfnamefont {I.}~\bibnamefont {Vela-P{\'{e}}rez}}, \bibinfo
  {author} {\bibfnamefont {D.}~\bibnamefont {Trabert}}, \bibinfo {author}
  {\bibfnamefont {S.}~\bibnamefont {Grundmann}}, \bibinfo {author}
  {\bibfnamefont {D.}~\bibnamefont {Tsitsonis}}, \bibinfo {author}
  {\bibfnamefont {J.}~\bibnamefont {Siebert}}, \bibinfo {author} {\bibfnamefont
  {A.}~\bibnamefont {Geyer}}, \bibinfo {author} {\bibfnamefont
  {N.}~\bibnamefont {Melzer}}, \bibinfo {author} {\bibfnamefont
  {C.}~\bibnamefont {Schwarz}}, \bibinfo {author} {\bibfnamefont
  {N.}~\bibnamefont {Anders}}, \bibinfo {author} {\bibfnamefont
  {L.}~\bibnamefont {Kaiser}}, \bibinfo {author} {\bibfnamefont
  {K.}~\bibnamefont {Fehre}}, \bibinfo {author} {\bibfnamefont
  {A.}~\bibnamefont {Hartung}}, \bibinfo {author} {\bibfnamefont
  {S.}~\bibnamefont {Eckart}}, \bibinfo {author} {\bibnamefont {{{L. Ph. H.
  Schmidt}}}}, \bibinfo {author} {\bibfnamefont {M.~S.}\ \bibnamefont
  {Sch{\"{o}}ffler}}, \bibinfo {author} {\bibfnamefont {V.~T.}\ \bibnamefont
  {Davis}}, \bibinfo {author} {\bibfnamefont {J.~B.}\ \bibnamefont {Williams}},
  \bibinfo {author} {\bibfnamefont {F.}~\bibnamefont {Trinter}}, \bibinfo
  {author} {\bibfnamefont {R.}~\bibnamefont {D{\"{o}}rner}}, \bibinfo {author}
  {\bibfnamefont {P.~V.}\ \bibnamefont {Demekhin}},\ and\ \bibinfo {author}
  {\bibfnamefont {T.}~\bibnamefont {Jahnke}},\ }\bibfield  {title} {\bibinfo
  {title} {{Measuring the photoelectron emission delay in the molecular
  frame}},\ }\href {https://doi.org/10.1038/s41467-021-26994-2} {\bibfield
  {journal} {\bibinfo  {journal} {Nat. Commun.}\ }\textbf {\bibinfo {volume}
  {12}},\ \bibinfo {pages} {6657} (\bibinfo {year} {2021})}\BibitemShut
  {NoStop}%
\bibitem [{\citenamefont {Holzmeier}\ \emph {et~al.}(2021)\citenamefont
  {Holzmeier}, \citenamefont {Joseph}, \citenamefont {Houver}, \citenamefont
  {Lebech}, \citenamefont {Dowek},\ and\ \citenamefont
  {Lucchese}}]{Holzmeier2021}%
  \BibitemOpen
  \bibfield  {author} {\bibinfo {author} {\bibfnamefont {F.}~\bibnamefont
  {Holzmeier}}, \bibinfo {author} {\bibfnamefont {J.}~\bibnamefont {Joseph}},
  \bibinfo {author} {\bibfnamefont {J.~C.}\ \bibnamefont {Houver}}, \bibinfo
  {author} {\bibfnamefont {M.}~\bibnamefont {Lebech}}, \bibinfo {author}
  {\bibfnamefont {D.}~\bibnamefont {Dowek}},\ and\ \bibinfo {author}
  {\bibfnamefont {R.~R.}\ \bibnamefont {Lucchese}},\ }\bibfield  {title}
  {\bibinfo {title} {{Influence of shape resonances on the angular dependence
  of molecular photoionization delays}},\ }\href
  {https://doi.org/10.1038/s41467-021-27360-y} {\bibfield  {journal} {\bibinfo
  {journal} {Nat. Commun.}\ }\textbf {\bibinfo {volume} {12}},\ \bibinfo
  {pages} {7343} (\bibinfo {year} {2021})}\BibitemShut {NoStop}%
\bibitem [{\citenamefont {Han}\ \emph {et~al.}(2018)\citenamefont {Han},
  \citenamefont {Ge}, \citenamefont {Shao}, \citenamefont {Gong},\ and\
  \citenamefont {Liu}}]{Han2018}%
  \BibitemOpen
  \bibfield  {author} {\bibinfo {author} {\bibfnamefont {M.}~\bibnamefont
  {Han}}, \bibinfo {author} {\bibfnamefont {P.}~\bibnamefont {Ge}}, \bibinfo
  {author} {\bibfnamefont {Y.}~\bibnamefont {Shao}}, \bibinfo {author}
  {\bibfnamefont {Q.}~\bibnamefont {Gong}},\ and\ \bibinfo {author}
  {\bibfnamefont {Y.}~\bibnamefont {Liu}},\ }\bibfield  {title} {\bibinfo
  {title} {{Attoclock Photoelectron Interferometry with Two-Color Corotating
  Circular Fields to Probe the Phase and the Amplitude of Emitting Wave
  Packets}},\ }\href {https://doi.org/10.1103/PhysRevLett.120.073202}
  {\bibfield  {journal} {\bibinfo  {journal} {Phys. Rev. Lett.}\ }\textbf
  {\bibinfo {volume} {120}},\ \bibinfo {pages} {073202} (\bibinfo {year}
  {2018})}\BibitemShut {NoStop}%
\bibitem [{\citenamefont {Eckart}(2020)}]{Eckart_HASEtheo}%
  \BibitemOpen
  \bibfield  {author} {\bibinfo {author} {\bibfnamefont {S.}~\bibnamefont
  {Eckart}},\ }\bibfield  {title} {\bibinfo {title} {Holographic angular
  streaking of electrons and the {{Wigner}} time delay},\ }\href
  {https://doi.org/10.1103/PhysRevResearch.2.033248} {\bibfield  {journal}
  {\bibinfo  {journal} {Phys. Rev. Res.}\ }\textbf {\bibinfo {volume} {2}},\
  \bibinfo {pages} {033248} (\bibinfo {year} {2020})}\BibitemShut {NoStop}%
\bibitem [{\citenamefont {Trabert}\ \emph {et~al.}(2021)\citenamefont
  {Trabert}, \citenamefont {Brennecke}, \citenamefont {Fehre}, \citenamefont
  {Anders}, \citenamefont {Geyer}, \citenamefont {Grundmann}, \citenamefont
  {Sch{\"{o}}ffler}, \citenamefont {{{L. Ph. H. Schmidt}}}, \citenamefont
  {Jahnke}, \citenamefont {D{\"{o}}rner}, \citenamefont {Kunitski},\ and\
  \citenamefont {Eckart}}]{Trabert2021}%
  \BibitemOpen
  \bibfield  {author} {\bibinfo {author} {\bibfnamefont {D.}~\bibnamefont
  {Trabert}}, \bibinfo {author} {\bibfnamefont {S.}~\bibnamefont {Brennecke}},
  \bibinfo {author} {\bibfnamefont {K.}~\bibnamefont {Fehre}}, \bibinfo
  {author} {\bibfnamefont {N.}~\bibnamefont {Anders}}, \bibinfo {author}
  {\bibfnamefont {A.}~\bibnamefont {Geyer}}, \bibinfo {author} {\bibfnamefont
  {S.}~\bibnamefont {Grundmann}}, \bibinfo {author} {\bibfnamefont {M.~S.}\
  \bibnamefont {Sch{\"{o}}ffler}}, \bibinfo {author} {\bibnamefont {{{L. Ph. H.
  Schmidt}}}}, \bibinfo {author} {\bibfnamefont {T.}~\bibnamefont {Jahnke}},
  \bibinfo {author} {\bibfnamefont {R.}~\bibnamefont {D{\"{o}}rner}}, \bibinfo
  {author} {\bibfnamefont {M.}~\bibnamefont {Kunitski}},\ and\ \bibinfo
  {author} {\bibfnamefont {S.}~\bibnamefont {Eckart}},\ }\bibfield  {title}
  {\bibinfo {title} {{Angular dependence of the {{Wigner}} time delay upon
  tunnel ionization of H$_{2}$}},\ }\href
  {https://doi.org/10.1038/s41467-021-21845-6} {\bibfield  {journal} {\bibinfo
  {journal} {Nat. Commun.}\ }\textbf {\bibinfo {volume} {12}},\ \bibinfo
  {pages} {1697} (\bibinfo {year} {2021})}\BibitemShut {NoStop}%
\bibitem [{\citenamefont {Trabert}\ \emph
  {et~al.}(2023{\natexlab{a}})\citenamefont {Trabert}, \citenamefont {Anders},
  \citenamefont {Geyer}, \citenamefont {Hofmann}, \citenamefont {Sch\"offler},
  \citenamefont {{{L. Ph. H. Schmidt}}}, \citenamefont {Jahnke}, \citenamefont
  {Kunitski}, \citenamefont {D\"orner},\ and\ \citenamefont
  {Eckart}}]{ArgonHASE}%
  \BibitemOpen
  \bibfield  {author} {\bibinfo {author} {\bibfnamefont {D.}~\bibnamefont
  {Trabert}}, \bibinfo {author} {\bibfnamefont {N.}~\bibnamefont {Anders}},
  \bibinfo {author} {\bibfnamefont {A.}~\bibnamefont {Geyer}}, \bibinfo
  {author} {\bibfnamefont {M.}~\bibnamefont {Hofmann}}, \bibinfo {author}
  {\bibfnamefont {M.~S.}\ \bibnamefont {Sch\"offler}}, \bibinfo {author}
  {\bibnamefont {{{L. Ph. H. Schmidt}}}}, \bibinfo {author} {\bibfnamefont
  {T.}~\bibnamefont {Jahnke}}, \bibinfo {author} {\bibfnamefont
  {M.}~\bibnamefont {Kunitski}}, \bibinfo {author} {\bibfnamefont
  {R.}~\bibnamefont {D\"orner}},\ and\ \bibinfo {author} {\bibfnamefont
  {S.}~\bibnamefont {Eckart}},\ }\bibfield  {title} {\bibinfo {title} {Angular
  dependence of the {{Wigner}} time delay upon strong-field ionization from an
  aligned $p$ orbital},\ }\href
  {https://doi.org/10.1103/PhysRevResearch.5.023118} {\bibfield  {journal}
  {\bibinfo  {journal} {Phys. Rev. Res.}\ }\textbf {\bibinfo {volume} {5}},\
  \bibinfo {pages} {023118} (\bibinfo {year} {2023}{\natexlab{a}})}\BibitemShut
  {NoStop}%
\bibitem [{\citenamefont {Trabert}\ \emph
  {et~al.}(2023{\natexlab{b}})\citenamefont {Trabert}, \citenamefont {Geyer},
  \citenamefont {Anders}, \citenamefont {Hofmann}, \citenamefont {Sch\"offler},
  \citenamefont {{{L. Ph. H. Schmidt}}}, \citenamefont {Jahnke}, \citenamefont
  {Kunitski}, \citenamefont {D\"orner},\ and\ \citenamefont
  {Eckart}}]{Trabert2023}%
  \BibitemOpen
  \bibfield  {author} {\bibinfo {author} {\bibfnamefont {D.}~\bibnamefont
  {Trabert}}, \bibinfo {author} {\bibfnamefont {A.}~\bibnamefont {Geyer}},
  \bibinfo {author} {\bibfnamefont {N.}~\bibnamefont {Anders}}, \bibinfo
  {author} {\bibfnamefont {M.}~\bibnamefont {Hofmann}}, \bibinfo {author}
  {\bibfnamefont {M.~S.}\ \bibnamefont {Sch\"offler}}, \bibinfo {author}
  {\bibnamefont {{{L. Ph. H. Schmidt}}}}, \bibinfo {author} {\bibfnamefont
  {T.}~\bibnamefont {Jahnke}}, \bibinfo {author} {\bibfnamefont
  {M.}~\bibnamefont {Kunitski}}, \bibinfo {author} {\bibfnamefont
  {R.}~\bibnamefont {D\"orner}},\ and\ \bibinfo {author} {\bibfnamefont
  {S.}~\bibnamefont {Eckart}},\ }\bibfield  {title} {\bibinfo {title} {Ideal
  two-color field ratio for holographic angular streaking of electrons},\
  }\href {https://doi.org/10.1103/PhysRevResearch.5.043245} {\bibfield
  {journal} {\bibinfo  {journal} {Phys. Rev. Res.}\ }\textbf {\bibinfo {volume}
  {5}},\ \bibinfo {pages} {043245} (\bibinfo {year}
  {2023}{\natexlab{b}})}\BibitemShut {NoStop}%
\bibitem [{\citenamefont {Guo}\ \emph {et~al.}(2022)\citenamefont {Guo},
  \citenamefont {Ge}, \citenamefont {Fang}, \citenamefont {Dou}, \citenamefont
  {Yu}, \citenamefont {Wang}, \citenamefont {Gong},\ and\ \citenamefont
  {Liu}}]{Guo2022}%
  \BibitemOpen
  \bibfield  {author} {\bibinfo {author} {\bibfnamefont {Z.}~\bibnamefont
  {Guo}}, \bibinfo {author} {\bibfnamefont {P.}~\bibnamefont {Ge}}, \bibinfo
  {author} {\bibfnamefont {Y.}~\bibnamefont {Fang}}, \bibinfo {author}
  {\bibfnamefont {Y.}~\bibnamefont {Dou}}, \bibinfo {author} {\bibfnamefont
  {X.}~\bibnamefont {Yu}}, \bibinfo {author} {\bibfnamefont {J.}~\bibnamefont
  {Wang}}, \bibinfo {author} {\bibfnamefont {Q.}~\bibnamefont {Gong}},\ and\
  \bibinfo {author} {\bibfnamefont {Y.}~\bibnamefont {Liu}},\ }\bibfield
  {title} {\bibinfo {title} {Probing molecular frame {W}igner time delay and
  electron wavepacket phase structure of {CO} molecule},\ }\href
  {https://doi.org/10.34133/2022/9802917} {\bibfield  {journal} {\bibinfo
  {journal} {Ultrafast Science}\ }\textbf {\bibinfo {volume} {2022}},\ \bibinfo
  {pages} {9802917} (\bibinfo {year} {2022})}\BibitemShut {NoStop}%
\bibitem [{\citenamefont {Eckart}\ \emph {et~al.}(2020)\citenamefont {Eckart},
  \citenamefont {Trabert}, \citenamefont {Fehre}, \citenamefont {Geyer},
  \citenamefont {Rist}, \citenamefont {Lin}, \citenamefont {Trinter},
  \citenamefont {{{L. Ph. H. Schmidt}}}, \citenamefont {Sch{\"{o}}ffler},
  \citenamefont {Jahnke}, \citenamefont {Kunitski},\ and\ \citenamefont
  {D{\"{o}}rner}}]{Eckart2020_sideband}%
  \BibitemOpen
  \bibfield  {author} {\bibinfo {author} {\bibfnamefont {S.}~\bibnamefont
  {Eckart}}, \bibinfo {author} {\bibfnamefont {D.}~\bibnamefont {Trabert}},
  \bibinfo {author} {\bibfnamefont {K.}~\bibnamefont {Fehre}}, \bibinfo
  {author} {\bibfnamefont {A.}~\bibnamefont {Geyer}}, \bibinfo {author}
  {\bibfnamefont {J.}~\bibnamefont {Rist}}, \bibinfo {author} {\bibfnamefont
  {K.}~\bibnamefont {Lin}}, \bibinfo {author} {\bibfnamefont {F.}~\bibnamefont
  {Trinter}}, \bibinfo {author} {\bibnamefont {{{L. Ph. H. Schmidt}}}},
  \bibinfo {author} {\bibfnamefont {M.~S.}\ \bibnamefont {Sch{\"{o}}ffler}},
  \bibinfo {author} {\bibfnamefont {T.}~\bibnamefont {Jahnke}}, \bibinfo
  {author} {\bibfnamefont {M.}~\bibnamefont {Kunitski}},\ and\ \bibinfo
  {author} {\bibfnamefont {R.}~\bibnamefont {D{\"{o}}rner}},\ }\bibfield
  {title} {\bibinfo {title} {{Sideband modulation by subcycle interference}},\
  }\href {https://doi.org/10.1103/PhysRevA.102.043115} {\bibfield  {journal}
  {\bibinfo  {journal} {Phys. Rev. A}\ }\textbf {\bibinfo {volume} {102}},\
  \bibinfo {pages} {043115} (\bibinfo {year} {2020})}\BibitemShut {NoStop}%
\bibitem [{\citenamefont {Murray}\ \emph {et~al.}(2010)\citenamefont {Murray},
  \citenamefont {Liu},\ and\ \citenamefont {Ivanov}}]{Murray2010}%
  \BibitemOpen
  \bibfield  {author} {\bibinfo {author} {\bibfnamefont {R.}~\bibnamefont
  {Murray}}, \bibinfo {author} {\bibfnamefont {W.-K.}\ \bibnamefont {Liu}},\
  and\ \bibinfo {author} {\bibfnamefont {M.~Y.}\ \bibnamefont {Ivanov}},\
  }\bibfield  {title} {\bibinfo {title} {{Partial Fourier-transform approach to
  tunnel ionization: Atomic systems}},\ }\href
  {https://doi.org/10.1103/PhysRevA.81.023413} {\bibfield  {journal} {\bibinfo
  {journal} {Phys. Rev. A}\ }\textbf {\bibinfo {volume} {81}},\ \bibinfo
  {pages} {023413} (\bibinfo {year} {2010})}\BibitemShut {NoStop}%
\bibitem [{\citenamefont {Liu}\ and\ \citenamefont {Liu}(2016)}]{Liu_2016}%
  \BibitemOpen
  \bibfield  {author} {\bibinfo {author} {\bibfnamefont {M.}~\bibnamefont
  {Liu}}\ and\ \bibinfo {author} {\bibfnamefont {Y.}~\bibnamefont {Liu}},\
  }\bibfield  {title} {\bibinfo {title} {{Application of the
  partial-Fourier-transform approach for tunnel ionization of molecules}},\
  }\href {https://doi.org/10.1103/PhysRevA.93.043426} {\bibfield  {journal}
  {\bibinfo  {journal} {Phys. Rev. A}\ }\textbf {\bibinfo {volume} {93}},\
  \bibinfo {pages} {043426} (\bibinfo {year} {2016})}\BibitemShut {NoStop}%
\bibitem [{Note1()}]{Note1}%
  \BibitemOpen
  \bibinfo {note} {Fig.~\ref {fig1}(a) shows a two-dimensional logarithmic
  density distribution $\vert {}\psi _r(x,y)\vert {}^2$ which is described by
  the asymptotic part of the $5\sigma $ highest occupied molecular orbital
  (HOMO) of CO in position space, modeled by $\psi ^\protect \mathrm
  {(m)}_\protect \mathrm {n}(\protect \mathbf {r})= \DOTSB \sum@ \slimits@ _{l}
  C_{lm} r^{(Z_{c}/\eta )-1} e^{-\eta r} Y_{lm}(\theta , \varphi )$ of the
  molecular Ammosov-Delone-Krainov (MO-ADK) model from Ref.~\cite {Zhao2010PRA}
  (with the ionization potential $I_\protect \mathrm {p} = 14.01$\protect
  \nobreak \protect \,eV, the bond length $R=1.128\protect \nobreak \protect
  \,\protect \text {\r A}$, $\eta =\protect \sqrt {2I_\protect \mathrm {p}}$,
  and the structure parameters $C_\protect \mathrm {00}=2.32$, $C_\protect
  \mathrm {10}=1.62$, $C_\protect \mathrm {20}=0.82$, $C_\protect \mathrm
  {30}=0.17$, and $C_\protect \mathrm {40}=0.05$).}\BibitemShut {Stop}%
\bibitem [{\citenamefont {Pfeiffer}\ \emph {et~al.}(2012)\citenamefont
  {Pfeiffer}, \citenamefont {Cirelli}, \citenamefont {Smolarski}, \citenamefont
  {Dimitrovski}, \citenamefont {Abu-samha}, \citenamefont {Madsen},\ and\
  \citenamefont {Keller}}]{Pfeiffer2012}%
  \BibitemOpen
  \bibfield  {author} {\bibinfo {author} {\bibfnamefont {A.~N.}\ \bibnamefont
  {Pfeiffer}}, \bibinfo {author} {\bibfnamefont {C.}~\bibnamefont {Cirelli}},
  \bibinfo {author} {\bibfnamefont {M.}~\bibnamefont {Smolarski}}, \bibinfo
  {author} {\bibfnamefont {D.}~\bibnamefont {Dimitrovski}}, \bibinfo {author}
  {\bibfnamefont {M.}~\bibnamefont {Abu-samha}}, \bibinfo {author}
  {\bibfnamefont {L.~B.}\ \bibnamefont {Madsen}},\ and\ \bibinfo {author}
  {\bibfnamefont {U.}~\bibnamefont {Keller}},\ }\bibfield  {title} {\bibinfo
  {title} {{Attoclock reveals natural coordinates of the laser-induced
  tunnelling current flow in atoms}},\ }\href
  {https://doi.org/10.1038/nphys2125} {\bibfield  {journal} {\bibinfo
  {journal} {Nat. Phys.}\ }\textbf {\bibinfo {volume} {8}},\ \bibinfo {pages}
  {76} (\bibinfo {year} {2012})}\BibitemShut {NoStop}%
\bibitem [{\citenamefont {Hofmann}\ \emph {et~al.}(2013)\citenamefont
  {Hofmann}, \citenamefont {Landsman}, \citenamefont {Cirelli}, \citenamefont
  {Pfeiffer},\ and\ \citenamefont {Keller}}]{Hofmann2013}%
  \BibitemOpen
  \bibfield  {author} {\bibinfo {author} {\bibfnamefont {C.}~\bibnamefont
  {Hofmann}}, \bibinfo {author} {\bibfnamefont {A.~S.}\ \bibnamefont
  {Landsman}}, \bibinfo {author} {\bibfnamefont {C.}~\bibnamefont {Cirelli}},
  \bibinfo {author} {\bibfnamefont {A.~N.}\ \bibnamefont {Pfeiffer}},\ and\
  \bibinfo {author} {\bibfnamefont {U.}~\bibnamefont {Keller}},\ }\bibfield
  {title} {\bibinfo {title} {{Comparison of different approaches to the
  longitudinal momentum spread after tunnel ionization}},\ }\href
  {https://doi.org/10.1088/0953-4075/46/12/125601} {\bibfield  {journal}
  {\bibinfo  {journal} {J. Phys. B At. Mol. Opt. Phys.}\ }\textbf {\bibinfo
  {volume} {46}},\ \bibinfo {pages} {125601} (\bibinfo {year}
  {2013})}\BibitemShut {NoStop}%
\bibitem [{\citenamefont {D{\"{o}}rner}\ \emph {et~al.}(2000)\citenamefont
  {D{\"{o}}rner}, \citenamefont {Mergel}, \citenamefont {Jagutzki},
  \citenamefont {Spielberger}, \citenamefont {Ullrich}, \citenamefont
  {Moshammer},\ and\ \citenamefont {Schmidt-B{\"{o}}cking}}]{Dorner2000}%
  \BibitemOpen
  \bibfield  {author} {\bibinfo {author} {\bibfnamefont {R.}~\bibnamefont
  {D{\"{o}}rner}}, \bibinfo {author} {\bibfnamefont {V.}~\bibnamefont
  {Mergel}}, \bibinfo {author} {\bibfnamefont {O.}~\bibnamefont {Jagutzki}},
  \bibinfo {author} {\bibfnamefont {L.}~\bibnamefont {Spielberger}}, \bibinfo
  {author} {\bibfnamefont {J.}~\bibnamefont {Ullrich}}, \bibinfo {author}
  {\bibfnamefont {R.}~\bibnamefont {Moshammer}},\ and\ \bibinfo {author}
  {\bibfnamefont {H.}~\bibnamefont {Schmidt-B{\"{o}}cking}},\ }\bibfield
  {title} {\bibinfo {title} {{Cold Target Recoil Ion Momentum Spectroscopy: a
  ‘momentum microscope' to view atomic collision dynamics}},\ }\href
  {https://doi.org/10.1016/S0370-1573(99)00109-X} {\bibfield  {journal}
  {\bibinfo  {journal} {Phys. Rep.}\ }\textbf {\bibinfo {volume} {330}},\
  \bibinfo {pages} {95} (\bibinfo {year} {2000})}\BibitemShut {NoStop}%
\bibitem [{\citenamefont {Ullrich}\ \emph {et~al.}(2003)\citenamefont
  {Ullrich}, \citenamefont {Moshammer}, \citenamefont {Dorn}, \citenamefont
  {D{\"o}rner}, \citenamefont {{{L. Ph. H. Schmidt}}},\ and\ \citenamefont
  {Schmidt-B{\"o}cking}}]{Ullrich_2003}%
  \BibitemOpen
  \bibfield  {author} {\bibinfo {author} {\bibfnamefont {J.}~\bibnamefont
  {Ullrich}}, \bibinfo {author} {\bibfnamefont {R.}~\bibnamefont {Moshammer}},
  \bibinfo {author} {\bibfnamefont {A.}~\bibnamefont {Dorn}}, \bibinfo {author}
  {\bibfnamefont {R.}~\bibnamefont {D{\"o}rner}}, \bibinfo {author}
  {\bibnamefont {{{L. Ph. H. Schmidt}}}},\ and\ \bibinfo {author}
  {\bibfnamefont {H.}~\bibnamefont {Schmidt-B{\"o}cking}},\ }\bibfield  {title}
  {\bibinfo {title} {Recoil-ion and electron momentum spectroscopy:
  reaction-microscopes},\ }\href@noop {} {\bibfield  {journal} {\bibinfo
  {journal} {Rep. Prog. Phys.}\ }\textbf {\bibinfo {volume} {66}},\ \bibinfo
  {pages} {1463} (\bibinfo {year} {2003})}\BibitemShut {NoStop}%
\bibitem [{\citenamefont {Jagutzki}\ \emph {et~al.}(2002)\citenamefont
  {Jagutzki}, \citenamefont {Cerezo}, \citenamefont {Czasch}, \citenamefont
  {D{\"o}rner}, \citenamefont {Hattas}, \citenamefont {Huang}, \citenamefont
  {Mergel}, \citenamefont {Spillmann}, \citenamefont {Ullmann-Pfleger},
  \citenamefont {Weber}, \citenamefont {Schmidt-B{\"o}cking},\ and\
  \citenamefont {Smith}}]{Jagutzki_2002}%
  \BibitemOpen
  \bibfield  {author} {\bibinfo {author} {\bibfnamefont {O.}~\bibnamefont
  {Jagutzki}}, \bibinfo {author} {\bibfnamefont {A.}~\bibnamefont {Cerezo}},
  \bibinfo {author} {\bibfnamefont {A.}~\bibnamefont {Czasch}}, \bibinfo
  {author} {\bibfnamefont {R.}~\bibnamefont {D{\"o}rner}}, \bibinfo {author}
  {\bibfnamefont {M.}~\bibnamefont {Hattas}}, \bibinfo {author} {\bibfnamefont
  {M.}~\bibnamefont {Huang}}, \bibinfo {author} {\bibfnamefont
  {V.}~\bibnamefont {Mergel}}, \bibinfo {author} {\bibfnamefont
  {U.}~\bibnamefont {Spillmann}}, \bibinfo {author} {\bibfnamefont
  {K.}~\bibnamefont {Ullmann-Pfleger}}, \bibinfo {author} {\bibfnamefont
  {T.}~\bibnamefont {Weber}}, \bibinfo {author} {\bibfnamefont
  {H.}~\bibnamefont {Schmidt-B{\"o}cking}},\ and\ \bibinfo {author}
  {\bibfnamefont {G.~D.~W.}\ \bibnamefont {Smith}},\ }\bibfield  {title}
  {\bibinfo {title} {Multiple hit readout of a microchannel plate detector with
  a three-layer delay-line anode},\ }\href@noop {} {\bibfield  {journal}
  {\bibinfo  {journal} {IEEE Trans. Nucl. Sci.}\ }\textbf {\bibinfo {volume}
  {49}},\ \bibinfo {pages} {2477} (\bibinfo {year} {2002})}\BibitemShut
  {NoStop}%
\bibitem [{\citenamefont {Eckart}\ \emph {et~al.}(2018)\citenamefont {Eckart},
  \citenamefont {Fehre}, \citenamefont {Eicke}, \citenamefont {Hartung},
  \citenamefont {Rist}, \citenamefont {Trabert}, \citenamefont {Strenger},
  \citenamefont {Pier}, \citenamefont {{{L. Ph. H. Schmidt}}}, \citenamefont
  {Jahnke}, \citenamefont {Sch\"offler}, \citenamefont {Lein}, \citenamefont
  {Kunitski},\ and\ \citenamefont {D\"orner}}]{Eckart2018_Offsets}%
  \BibitemOpen
  \bibfield  {author} {\bibinfo {author} {\bibfnamefont {S.}~\bibnamefont
  {Eckart}}, \bibinfo {author} {\bibfnamefont {K.}~\bibnamefont {Fehre}},
  \bibinfo {author} {\bibfnamefont {N.}~\bibnamefont {Eicke}}, \bibinfo
  {author} {\bibfnamefont {A.}~\bibnamefont {Hartung}}, \bibinfo {author}
  {\bibfnamefont {J.}~\bibnamefont {Rist}}, \bibinfo {author} {\bibfnamefont
  {D.}~\bibnamefont {Trabert}}, \bibinfo {author} {\bibfnamefont
  {N.}~\bibnamefont {Strenger}}, \bibinfo {author} {\bibfnamefont
  {A.}~\bibnamefont {Pier}}, \bibinfo {author} {\bibnamefont {{{L. Ph. H.
  Schmidt}}}}, \bibinfo {author} {\bibfnamefont {T.}~\bibnamefont {Jahnke}},
  \bibinfo {author} {\bibfnamefont {M.~S.}\ \bibnamefont {Sch\"offler}},
  \bibinfo {author} {\bibfnamefont {M.}~\bibnamefont {Lein}}, \bibinfo {author}
  {\bibfnamefont {M.}~\bibnamefont {Kunitski}},\ and\ \bibinfo {author}
  {\bibfnamefont {R.}~\bibnamefont {D\"orner}},\ }\bibfield  {title} {\bibinfo
  {title} {Direct experimental access to the nonadiabatic initial momentum
  offset upon tunnel ionization},\ }\href
  {https://doi.org/10.1103/PhysRevLett.121.163202} {\bibfield  {journal}
  {\bibinfo  {journal} {Phys. Rev. Lett.}\ }\textbf {\bibinfo {volume} {121}},\
  \bibinfo {pages} {163202} (\bibinfo {year} {2018})}\BibitemShut {NoStop}%
\bibitem [{\citenamefont {Eckart}\ \emph {et~al.}(2016)\citenamefont {Eckart},
  \citenamefont {Richter}, \citenamefont {Kunitski}, \citenamefont {Hartung},
  \citenamefont {Rist}, \citenamefont {Henrichs}, \citenamefont {Schlott},
  \citenamefont {Kang}, \citenamefont {Bauer}, \citenamefont {Sann},
  \citenamefont {{{L. Ph. H. Schmidt}}}, \citenamefont {Sch{\"{o}}ffler},
  \citenamefont {Jahnke},\ and\ \citenamefont {D{\"{o}}rner}}]{Eckart2016}%
  \BibitemOpen
  \bibfield  {author} {\bibinfo {author} {\bibfnamefont {S.}~\bibnamefont
  {Eckart}}, \bibinfo {author} {\bibfnamefont {M.}~\bibnamefont {Richter}},
  \bibinfo {author} {\bibfnamefont {M.}~\bibnamefont {Kunitski}}, \bibinfo
  {author} {\bibfnamefont {A.}~\bibnamefont {Hartung}}, \bibinfo {author}
  {\bibfnamefont {J.}~\bibnamefont {Rist}}, \bibinfo {author} {\bibfnamefont
  {K.}~\bibnamefont {Henrichs}}, \bibinfo {author} {\bibfnamefont
  {N.}~\bibnamefont {Schlott}}, \bibinfo {author} {\bibfnamefont
  {H.}~\bibnamefont {Kang}}, \bibinfo {author} {\bibfnamefont {T.}~\bibnamefont
  {Bauer}}, \bibinfo {author} {\bibfnamefont {H.}~\bibnamefont {Sann}},
  \bibinfo {author} {\bibnamefont {{{L. Ph. H. Schmidt}}}}, \bibinfo {author}
  {\bibfnamefont {M.}~\bibnamefont {Sch{\"{o}}ffler}}, \bibinfo {author}
  {\bibfnamefont {T.}~\bibnamefont {Jahnke}},\ and\ \bibinfo {author}
  {\bibfnamefont {R.}~\bibnamefont {D{\"{o}}rner}},\ }\bibfield  {title}
  {\bibinfo {title} {{Nonsequential Double Ionization by Counterrotating
  Circularly Polarized Two-Color Laser Fields}},\ }\href
  {https://doi.org/10.1103/PhysRevLett.117.133202} {\bibfield  {journal}
  {\bibinfo  {journal} {Phys. Rev. Lett.}\ }\textbf {\bibinfo {volume} {117}},\
  \bibinfo {pages} {133202} (\bibinfo {year} {2016})}\BibitemShut {NoStop}%
\bibitem [{\citenamefont {Eckle}\ \emph {et~al.}(2008)\citenamefont {Eckle},
  \citenamefont {Pfeiffer}, \citenamefont {Cirelli}, \citenamefont {Staudte},
  \citenamefont {D{\"{o}}rner}, \citenamefont {Muller}, \citenamefont
  {B{\"{u}}ttiker},\ and\ \citenamefont {Keller}}]{Eckle2008}%
  \BibitemOpen
  \bibfield  {author} {\bibinfo {author} {\bibfnamefont {P.}~\bibnamefont
  {Eckle}}, \bibinfo {author} {\bibfnamefont {A.~N.}\ \bibnamefont {Pfeiffer}},
  \bibinfo {author} {\bibfnamefont {C.}~\bibnamefont {Cirelli}}, \bibinfo
  {author} {\bibfnamefont {A.}~\bibnamefont {Staudte}}, \bibinfo {author}
  {\bibfnamefont {R.}~\bibnamefont {D{\"{o}}rner}}, \bibinfo {author}
  {\bibfnamefont {H.~G.}\ \bibnamefont {Muller}}, \bibinfo {author}
  {\bibfnamefont {M.}~\bibnamefont {B{\"{u}}ttiker}},\ and\ \bibinfo {author}
  {\bibfnamefont {U.}~\bibnamefont {Keller}},\ }\bibfield  {title} {\bibinfo
  {title} {{Attosecond Ionization and Tunneling Delay Time Measurements in
  Helium}},\ }\href {https://doi.org/10.1126/science.1163439} {\bibfield
  {journal} {\bibinfo  {journal} {Science}\ }\textbf {\bibinfo {volume}
  {322}},\ \bibinfo {pages} {1525} (\bibinfo {year} {2008})}\BibitemShut
  {NoStop}%
\bibitem [{\citenamefont {Zhao}\ \emph {et~al.}(2010)\citenamefont {Zhao},
  \citenamefont {Jin}, \citenamefont {Le}, \citenamefont {Jiang},\ and\
  \citenamefont {Lin}}]{Zhao2010PRA}%
  \BibitemOpen
  \bibfield  {author} {\bibinfo {author} {\bibfnamefont {S.-F.}\ \bibnamefont
  {Zhao}}, \bibinfo {author} {\bibfnamefont {C.}~\bibnamefont {Jin}}, \bibinfo
  {author} {\bibfnamefont {A.-T.}\ \bibnamefont {Le}}, \bibinfo {author}
  {\bibfnamefont {T.~F.}\ \bibnamefont {Jiang}},\ and\ \bibinfo {author}
  {\bibfnamefont {C.~D.}\ \bibnamefont {Lin}},\ }\bibfield  {title} {\bibinfo
  {title} {Determination of structure parameters in strong-field tunneling
  ionization theory of molecules},\ }\href
  {https://doi.org/10.1103/PhysRevA.81.033423} {\bibfield  {journal} {\bibinfo
  {journal} {Phys. Rev. A}\ }\textbf {\bibinfo {volume} {81}},\ \bibinfo
  {pages} {033423} (\bibinfo {year} {2010})}\BibitemShut {NoStop}%
\bibitem [{Note2()}]{Note2}%
  \BibitemOpen
  \bibinfo {note} {The simulated momentum mapping can be parameterized as
  $p_\protect \mathrm {init}(p_\protect \mathrm {elec}) = \protect \sqrt
  {p_\protect \mathrm {elec}^2 + b^2} - c$ with $b = 0.36\protect \,\protect
  \mathrm {a.u.}$ and $c = 0.275\protect \,\protect \mathrm
  {a.u.}$}\BibitemShut {NoStop}%
\bibitem [{\citenamefont {Wu}\ \emph {et~al.}(2012)\citenamefont {Wu},
  \citenamefont {{{L. Ph. H. Schmidt}}}, \citenamefont {Kunitski},
  \citenamefont {Meckel}, \citenamefont {Voss}, \citenamefont {Sann},
  \citenamefont {Kim}, \citenamefont {Jahnke}, \citenamefont {Czasch},\ and\
  \citenamefont {Dörner}}]{Wu2012}%
  \BibitemOpen
  \bibfield  {author} {\bibinfo {author} {\bibfnamefont {J.}~\bibnamefont
  {Wu}}, \bibinfo {author} {\bibnamefont {{{L. Ph. H. Schmidt}}}}, \bibinfo
  {author} {\bibfnamefont {M.}~\bibnamefont {Kunitski}}, \bibinfo {author}
  {\bibfnamefont {M.}~\bibnamefont {Meckel}}, \bibinfo {author} {\bibfnamefont
  {S.}~\bibnamefont {Voss}}, \bibinfo {author} {\bibfnamefont {H.}~\bibnamefont
  {Sann}}, \bibinfo {author} {\bibfnamefont {H.}~\bibnamefont {Kim}}, \bibinfo
  {author} {\bibfnamefont {T.}~\bibnamefont {Jahnke}}, \bibinfo {author}
  {\bibfnamefont {A.}~\bibnamefont {Czasch}},\ and\ \bibinfo {author}
  {\bibfnamefont {R.}~\bibnamefont {Dörner}},\ }\bibfield  {title} {\bibinfo
  {title} {Multiorbital tunneling ionization of the {CO} molecule},\ }\href
  {https://doi.org/10.1103/PhysRevLett.108.183001} {\bibfield  {journal}
  {\bibinfo  {journal} {Phys. Rev. Lett.}\ }\textbf {\bibinfo {volume} {108}},\
  \bibinfo {pages} {183001} (\bibinfo {year} {2012})}\BibitemShut {NoStop}%
\bibitem [{\citenamefont {Ivanov}\ \emph {et~al.}(2005)\citenamefont {Ivanov},
  \citenamefont {Spanner},\ and\ \citenamefont {Smirnova}}]{Misha2005}%
  \BibitemOpen
  \bibfield  {author} {\bibinfo {author} {\bibfnamefont {M.~Y.}\ \bibnamefont
  {Ivanov}}, \bibinfo {author} {\bibfnamefont {M.}~\bibnamefont {Spanner}},\
  and\ \bibinfo {author} {\bibfnamefont {O.}~\bibnamefont {Smirnova}},\
  }\bibfield  {title} {\bibinfo {title} {Anatomy of strong field ionization},\
  }\href {https://doi.org/10.1080/0950034042000275360} {\bibfield  {journal}
  {\bibinfo  {journal} {J. Mod. Opt.}\ }\textbf {\bibinfo {volume} {52}},\
  \bibinfo {pages} {165} (\bibinfo {year} {2005})}\BibitemShut {NoStop}%
\bibitem [{\citenamefont {Barth}\ and\ \citenamefont
  {Smirnova}(2011)}]{Barth2011}%
  \BibitemOpen
  \bibfield  {author} {\bibinfo {author} {\bibfnamefont {I.}~\bibnamefont
  {Barth}}\ and\ \bibinfo {author} {\bibfnamefont {O.}~\bibnamefont
  {Smirnova}},\ }\bibfield  {title} {\bibinfo {title} {{Nonadiabatic tunneling
  in circularly polarized laser fields: Physical picture and calculations}},\
  }\href {https://doi.org/10.1103/PhysRevA.84.063415} {\bibfield  {journal}
  {\bibinfo  {journal} {Phys. Rev. A}\ }\textbf {\bibinfo {volume} {84}},\
  \bibinfo {pages} {063415} (\bibinfo {year} {2011})}\BibitemShut {NoStop}%
\bibitem [{\citenamefont {Haertelt}\ \emph {et~al.}(2026)\citenamefont
  {Haertelt}, \citenamefont {Wu}, \citenamefont {Lai}, \citenamefont {Naumov},
  \citenamefont {Liu}, \citenamefont {Corkum},\ and\ \citenamefont
  {Staudte}}]{Haertelt2026}%
  \BibitemOpen
  \bibfield  {author} {\bibinfo {author} {\bibfnamefont {M.}~\bibnamefont
  {Haertelt}}, \bibinfo {author} {\bibfnamefont {W.}~\bibnamefont {Wu}},
  \bibinfo {author} {\bibfnamefont {X.}~\bibnamefont {Lai}}, \bibinfo {author}
  {\bibfnamefont {A.~Y.}\ \bibnamefont {Naumov}}, \bibinfo {author}
  {\bibfnamefont {X.}~\bibnamefont {Liu}}, \bibinfo {author} {\bibfnamefont
  {P.~B.}\ \bibnamefont {Corkum}},\ and\ \bibinfo {author} {\bibfnamefont
  {A.}~\bibnamefont {Staudte}},\ }\bibfield  {title} {\bibinfo {title}
  {Photoelectron holography of a heteronuclear molecule},\ }\href
  {https://doi.org/10.1103/z4xn-m554} {\bibfield  {journal} {\bibinfo
  {journal} {Phys. Rev. Lett.}\ }\textbf {\bibinfo {volume} {136}},\ \bibinfo
  {pages} {153202} (\bibinfo {year} {2026})}\BibitemShut {NoStop}%
\end{thebibliography}%

\end{document}